\documentclass[fleqn,10pt]{wlscirep}
\usepackage{amsmath,amssymb}
\usepackage{amsmath, amsthm, amssymb}
\usepackage{graphicx}
\usepackage{epstopdf}

\title{Rectifying the output of vibrational piezoelectric energy harvester using quantum dots}
\author[1,*]{Lijie Li}

\affil[1]{Multidisciplinary Nanotechnology Centre, College of Engineering, Swansea University, Bay Campus, Swansea, SA1 8EN, UK}

\affil[*]{L.Li@swansea.ac.uk}

\begin{abstract}
 Piezoelectric energy harvester scavenges mechanical vibrations and generates electricity. Researchers have strived to optimize the electromechanical structures and to design necessary external power management circuits, aiming to deliver high power and rectified outputs ready for serving as batteries. Complex deformation of the mechanical structure results in charges with opposite polarities appearing on same surface, leading to current loss in the attached metal electrode. External power management circuits such as rectifiers comprise diodes that consume power and have undesirable forward bias. To address the above issues, we devise a novel integrated piezoelectric energy harvesting device that is structured by stacking a layer of quantum dots (QDs) and a layer of piezoelectric material. We find that the QD can rectify electrical charges generated from the piezoelectric material because of its adaptable conductance to the electrochemical potentials of both sides of the QDs layer, so that electrical current causing energy loss on the same surface of the piezoelectric material can be minimized. The QDs layer has the potential to replace external rectification circuits providing a much more compact and less power-consumption solution. 
\end{abstract}


\begin{document}

\maketitle

\section*{Introduction}
Piezoelectric effect is exhibited in crystals with no centre of symmetry. Under an external stress, a net polarization appears on the surface of the material, generating piezoelectric voltage, inducing electrical charges. Mechanical-to-electrical energy harvesting devices based on piezoelectric materials have been receiving much attention in recent decades. \cite{M2E1} \cite{M2E2} \cite{M2E3} Moreover it has been successfully advanced to nanoscale energy devices recently. \cite{zlwang} \cite{zlwangnn} Note that charge can only be continuously generated when there is a change in the applied force, which means that a static force only generates a fixed amount charges that will leak away quickly across the external circuit (internal screening effect is neglected). The time taken for leaking away all generated charges depends on the input impedance of the external circuit. For instance a 1$k\Omega$ load resistance $R_L$ and a 1$nF$ internal capacitance $C_P$ of the piezoelectric transducer results in a time constant of $ R_L \times C_P =1 \mu s $, which implies that after 1 $\mu s$ all generated charges from a static mechanical force disappear. For this reason the device works mainly under AC conditions for energy harvesting purpose, in order to continuously generate electrical charges. This brought out the concept of 'vibrational' energy harvesting structures. The optimal power generation is when the device operates at its resonant frequency \cite{M2E4}, which leads to essential additions of external rectifying circuits \cite{Circ1} \cite{Circ2} \cite{Circ3} , causing unavoidable power consumptions. The vibrational energy harvesters usually appear in the form of a cantilever with one end fixed, the other suspended \cite{Canti1} \cite{Canti2} \cite{Canti3} \cite{Canti4} , as mechanical cantilevers are able to generate larger strain, easy to fabricate, and having lower fundamental frequencies to match with ambient mechanical sources. Although tremendous efforts have been spent to increase the output power of the cantilever harvester, including some innovative structural designs. \cite{Wideband1} \cite{Wideband2} \cite{Li2} There is an ongoing issue of designing low power consumption AC-DC converters for rectifying the AC output of the harvester \cite{Circ4} \cite{Circ5} \cite{Circ6} . Complex vibrations of the cantilever subjecting to arbitrary excitations from ambient sources sometimes result in both positive and negative charges accumulating on the same surface, subsequently introducing electrical current on the metal electrode attaching to that piezoelectric surface. This amount of power loss is not compensatable using the external circuits. We used segmentation protocol aiming for collecting positive and negative charges separately \cite{Li1} \cite{Li3} \cite{Lipatent} , essentially creating many sub-cells. However it will require more complex external circuits to handle these sub-cells, and the consequence will be more electrical power dissipated in the power management circuits. Inspired from the electron quantum transport theory of conductors \cite{DattaBook}, we propose to coat a layer of QDs onto the piezoelectric cantilever (schematic diagram is shown in \textbf{Figure 1a}). The QD layer is able to function as an integrated filter or rectifier that has adaptable resistance under different chemical potentials of both sides of the QDs layer, which are the metal electrode and piezoelectric material respectively. Thus charges with different polarities and quantities cause variation of electrochemical potential of the piezoelectric surface attaching to the QDs layer, adjusting the resistance of the QDs layer. The theory and method unpinning this hypothesis is detailed in the next section, followed by the numerical case study. This new method is much more efficient than previous approaches of using external circuits or segmenting electrodes, since it does not need additional circuits consisting of active electronic components like transistors, diodes. This design seamlessly integrates power management circuits with the actual transducer, moreover it significantly reduces power loss on the same surface of the piezoelectric material in the cases of the harvester experiencing complex deformations.    

\section{Models and Methods}

The multiphysical treatment for this new concept is split into three steps, namely mechanical-to-electrical (MTE), charge-to-electrochemical potential (CTE), and elastic tunnelling (ET) through the quantum dot. Euler$-$Bernoulli beam theory together with the constitutive piezoelectric equation will be used in the MTE stage, CTE simulation is based on semiconductor physics, and ET simulation is built on the quantum mechanics theory. This open-loop multi-physics simulation links the external applied mechanical excitation to the conductance of the QD, through which the rectifying function of this concept is demonstrated. Finally this integrated device is connected to a load resistor for demonstration in the real applications.

\subsection{Mechanical strain to electrical charge}
To begin with the analysis, we describe the integrated energy harvester as the form of a cantilever, which consists of a layer of piezoelectric material acting as the strain to electrical charge converter, and a layer of quantum dots that can be spin coated or printed. In the real scenario, metal electrodes are plated on both sides of the composite cantilever. Due to very thin electrode layer, its mechanical contribution is neglected in the analysis. The length $L$, width $w$, and thickness $g$ of the piezoelectric layer are designated as 5 $mm$, 0.5 $mm$, and 50 $\mu m$ for the later numerical demonstration. AlN is chosen as the piezoelectric material with density $\rho$ of 3300 $kg/m^3$ \cite{AlNdensity}, Poisson's ratio $\nu$ and Young's modulus $Y$ of 350 $GPa$ \cite{AlNyoung}. The quantum dots layer is so thin (normally $< 1 \mu m$), and it can also be neglected in the mechanical analysis. The dynamic motion of the cantilever can be described using the mass-spring-damper equation \cite{Li4} 
\begin{equation}
m_t\ddot{z}+b\dot{z}+k_1z+k_2z^3=Acos(\omega t)
\end{equation} 
where $m_t$, $b$, $k_1$, $k_2$ are total effective mass $(\frac{33}{140}+100)Lwg\rho$ (here we attach an additional proof mass (100$\times Lwg\rho$) at the free end of the cantilever to reduce the resonant frequency), damping coefficient, linear stiffness, and nonlinear stiffness respectively. $z(t)$ is the tip displacement. The first derivative of the $z$ with respect to the time $t$ represents the velocity, and the second derivative denotes the acceleration. $A$ is the amplitude of the external periodic driving force, and $\omega$ is the driving frequency. When $\omega = \omega_0$ ($\omega_0$ being the natural frequency of the cantilever $\omega_0=\sqrt{\frac{k_1}{m_t}}$, here we only analyse the fundamental resonance), the deformation of the cantilever reaches to the maximum. Here we set the driving frequency same as the nature frequency of the cantilever to ensure an optimal match between the excitation and the beam. Under the external driving, the cantilever vibrates, the deflected cantilever can be treated as an arc. The curvature $1/r$ ($r$ is the radius) of this arc is expressed with respect to the tip deflection $z$
\begin{equation}
\frac{1}{r(t)}=\frac{2z(t)}{L^2}
\end{equation} 
The axial strain on the surface of the piezoelectric layer along the $x$-axis can be obtained in terms of curvature and tip deflection \cite{Li4}
\begin{equation}
\varepsilon_x(t)=\frac{g}{2r(t)}
\end{equation}
The axial stress $\sigma$ in the piezoelectric layer is derived through the stress-strain relation, that is $\sigma_x(t)=Y\varepsilon_x(t)$. The piezoelectric constitutive equations relate four field variables stress components $\Sigma$, strain components $S$, electric field components $E$, and the electric displacement components $D$, which can be described as $\begin{bmatrix} S \\ D \end{bmatrix} = \begin{bmatrix} s^E & d^t \\ d & \varepsilon^T \end{bmatrix} \begin{bmatrix} \Sigma \\ E \end{bmatrix}$. \cite{PiezoBook} \cite{Piezo} We shall focus on the generated electric displacement $D$ from the mechanical stress $\Sigma$. Assuming no external electric field is applied to the cantilever, $D=d \Sigma$, $d$ being the piezoelectric constant. The electrical charges generated from a tip deflection $z(t)$ can therefore be obtained 
\begin{equation}
Q(t)=wLd_{31}\sigma_x(t)=\frac{wd_{31}Yg}{L}z(t)
\end{equation} 
where $d_{31}$ is the piezoelectric constant, relating the strain in the '1' direction ($x$-direction) to a generated field along the '3' direction ($z$-direction). In the simulation, we take $d_{31}$=$-$1.9 pC/N \cite{AlN1}. A upward deflection ($z>0$) corresponds to a compressive stress on the top surface, therefore inducing negative charges.

\subsection{Charge to electrochemical potential}
After obtaining piezoelectrically generated charges, electrochemical potentials (top and bottom surface) of the piezoelectric layer can be derived from the carrier density equation \cite{AQMbook}, which is
\begin{align}
n&=\int_{0}^{\infty} D_3(\mathcal{E})f_0(\mathcal{E})d\mathcal{E}\notag\\
&=\frac{1}{2\pi^2}(\frac{2m^*}{\hbar^2})^{3/2}\int_{0}^{\infty} \frac{\mathcal{E}^{1/2}}{\exp{(\frac{\mathcal{E}-\mu}{k_BT})}+1} d\mathcal{E}
\end{align}
where $D_3(\mathcal{E})$ is the density of state for 3-dimensional bulk materials, which takes the form of $\frac{1}{2\pi^2}(\frac{2m^*}{\hbar^2})^{3/2}\mathcal{E}^{1/2}$ \cite{Sezbook}, where $\hbar$ is the reduced Planck's constant ($1.05 \times 10^{-34} Js$), $m^*$ is the effective mass of electron $m^*=0.33m_0$, \cite{AlNmass} where $m_0$ is the bare electron mass $9.11 \times 10^{-31} Kg$. $n$ is the carrier density, which is varying upon deflection of the piezoelectric cantilever. The additional carrier density is calculated by $\Delta n=Q(t)/(wLd_p)$, $d_p$ denotes the depletion thickness on the surface of the piezoelectric, which can be approximated according to the Poisson equation $\frac{d^2\phi (z)}{dz^2}=\frac{e\Delta n}{\epsilon_s} $. Here for simplicity it is taken as a constant 10 nm. $f_0(\mathcal{E})$ is the Fermi distribution of electrons $\frac{1}{\exp{(\frac{\mathcal{E}-\mu}{k_BT})}+1}$, where $\mathcal{E}$ is the electron energy, $\mu$ stands for the chemical potential, $k_B$ is the Boltzmann constant ($8.62 \times 10^{-5} eV/K^{-1}$), and $T$ denotes the temperature. Iteration method/self-consistent method is used to obtain $\mu$ through the numerical approach \cite{Li5}. In the work we calculate the Fermi distributions of electrons in conduction band $f_e$ and holes in valence band $f_h$, where the electrochemical potentials $\mu_e$ and $\mu_h$ are electrochemical potentials for electrons and holes respectively. The effective mass of heavy hole is taken as $3.68m_0$ \cite{AlNmass}. Focusing on the interface between QD and piezoelectric material, accumulated electrons due to piezoelectric effect causes electrochemical potential move to the conduction band minimum $\mu_c$. Accumulated holes at the interface tend to shift the chemical potential to the valence band maximum $\mu_v$. The middle of the bandgap $g_b$ is set to potential 0, $\mu_c=g_b/2$, $\mu_v=-g_b/2$. To solve the equation (5), the integral is calculated from conduction band minimum to a large value of $\mathcal{E}_m$ or from $-\mathcal{E}_m$ to the valence band maximum, $\mathcal{E}_m=100\times k_BT$ is large enough to arrive at an accurate result. Calculated chemical potential $\mu_e$ and $\mu_h$ are then mapped to the energy graph considering the bandgap $g_b$. For AlN, $g_b=6.2 eV$ \cite{AlN2}. The mapped electrochemical potential has been designated as the chemical potential of the piezoelectric interface $\mu_P=[\mu_e+g_b/2, (n_e>n_h)]\cup [-\mu_h-g_b/2, (n_e<n_h)]$. \textbf{Figure 1b} schematically displays the energy graph of the metal-QD-piezoelectric material interface. It is assumed that the piezoelectric material is intrinsic with the initial chemical potential in the middle of the bandgap, i.e. internal screening \cite{Screen1} of piezoelectric polarization is minimal. This assumption is reasonable as the AlN films exhibit very low carrier concentration in the real case.

\subsection{Elastic tunnelling}
In the normal operation conditions of the integrated piezoelectric vibrational energy harvesting device, surrounding temperature is usually room temperature. The energy cost to add an electron to the dot is determined by the dot's self-capacitance, that is $e^2/C_{\Sigma}$. Simple estimation of the self-capacitance is $C_\Sigma=e^2/(8\varepsilon_r \varepsilon_0 R)$ \cite{QDtrans} where $R$ is the radius of a 2D dot. For a typical GaAs dot with radius of 50 nm, $C_\Sigma$ is calculated to be 0.046 fF, resulting in $e^2/C_\Sigma=3.5 meV$, much smaller than thermal energy at room temperature 300K, $k_BT=25.9meV$. Hence Coulomb blockade effect can be ignored in this analysis. After $\mu_P$ is obtained, electrical current $I$ flowing through the quantum dot is derived using the Landauer formula \cite{DattaBook} \cite{QDTE}, that is
\begin{equation}
I=\frac{e}{\hbar \pi}\int_{-\infty}^{\infty} \Gamma (\mathcal{E})(f_T-f_P)d\mathcal{E}
\end{equation}
where $f_T$ and $f_P$ are electron Fermi distributions for the top electrode (note the QD layer is deposited on the top surface) and piezoelectric layer respectively, $f_T=f(\mathcal{E}-\mu_T,T)=\frac{1}{\exp((\mathcal{E}-\mu_T)/(k_BT))+1}$, $f_P=f(\mathcal{E}-\mu_P,T)=\frac{1}{\exp((\mathcal{E}-\mu_P)/(k_BT))+1}$. In the analysis, temperature $T$ remains constant 300 $K$. Here we shall calculate the $\mu_P$ according to the previous section, and $\mu_T$ is assumed to be 0, aligning with the chemical potential of the initial state of the device. Without external mechanical excitation, the initial chemical potential of the piezoelectric layer $\mu_{P}^0$ is 0. $\Gamma (\mathcal{E})$ represents the transmission function of the two interfaces for each incident electron energy $\mathcal{E}$. It is governed by the Lorentzian equation
\begin{equation}
\Gamma (\mathcal{E})=\frac{\gamma_T\gamma_P}{(\mathcal{E}-\mathcal{E}_{qd})^2+(\frac{\gamma_T+\gamma_P}{2})^2}
\end{equation}
where the $\gamma_{T,P}$ represents the elastic tunnel coupling coefficients through electrode-quantum dot, and quantum dot-piezoelectric layer respectively. They are determined by $\gamma_0\exp(-\delta z/\xi)$  \cite{Li6}, here $\delta z$ denotes the distance between quantum dot edge and the electrode, quantum dot edge and piezoelectric layer. $\xi$ is the characteristic length of the dot defining the exponential decay for the probability of finding electron outside the dot, i.e. decay of the wavefunction of the electron, which can normally be calculated from theory of electron passing barriers. Here in this analysis it is assumed that $\gamma_0=k_BT$. For the simulation we set $\delta z=20 nm$, $\xi=2 nm$, then we have $\gamma_{T,P}=0.45 \times 10^{-4}k_BT$. It is worth noted that $\gamma_{T,P}$ can be adjusted by changing positions of the QDs to achieve various rectification effects. The energy level of the quantum dot $\mathcal{E}_{qd}$ is set to 1.1 $eV$ (Here the fluctuation of the $\mathcal{E}_{qd}$ due to electrons flowing through the dot is not considered). Here we design the in-plane spacing between dots to be much larger than $\xi$, so that the current flow between dots can be neglected. The conductance of the quantum dots layer is
\begin{equation}
G=\frac{I}{V_D}=\frac{eI}{\delta \mu}=\frac{e^2}{\hbar \pi \delta \mu}\int_{-\infty}^{\infty} \Gamma (\mathcal{E})(f_T-f_P)d\mathcal{E}
\end{equation}  
where the integral is calculated for the energy range of ($-200k_BT, 200k_BT$), which is sufficient to cover the varying potential of the piezoelectric material. Once $G$ is obtained, further analysis on the integrated harvester connecting with an external load is discussed. The complete circuit consists of a load resistor and the integrated QD-piezoelectric energy harvester (\textbf{Figure 1c}). It is assumed that the piezoelectric material has infinite resistance i.e. a pure capacitor. As the cantilever is vibrating under the external load, conductance/resistance of the quantum dots layer changes with the varying $\mu_P$, only allowing electrons having energy within a certain band passing through. The next section demonstrates numerical analysis of the integrated harvester.

\section{Results and Discussion}

A case study to quantitatively demonstrate this idea is detailed in this section. The numerical procedure begins with the mechanical analysis. In solving the equation (1), the amplitude of periodical driving force $A$ is set to $m_ta$ where $m_t$ is the sum of the cantilever mass and the proof mass, $a$ denotes the externally applied acceleration (10 $m/s^2$). The linear stiffness of the cantilever can be calculated from $k_1=(Ywg^3)/(4L^3)$ \cite{mechan1}, nonlinear stiffness $k_2=2k_1$, and the damping factor is usually very small for miniaturized devices. Here we use $b=2(k_1/m_t)^{1/2} \times 10^{-5}$. From the numerical solution of the equation (1), substitute tip deflection $z(t)$ into the equations (2)-(4), generated charge $Q(t)$ is calculated with respect to time $t$. \textbf{Figure 2} depicts the result of $z(t)$ and $Q(t)$. It is seen that the cantilever oscillates periodically with a maximum peak displacement of $1.1 \mu m$, generated charge oscillating in a similar manner with the peak amplitude of $3.66 \times 10^{-12} C$. Next the resulting charge values are input to the equation (5) to arrive at chemical potentials $\mu_e$ and $\mu_h$, which are mapped to the band diagram and shown together with $\mu_c$ and $\mu_v$ in \textbf{Figure 3}. It is seen from \textbf{Figure 3a} that in the low density range, a small addition of the carrier density causes chemical potential to deviate to $\mu_c$ or $\mu_v$ rapidly. But the chemical potential remains steady in the large density range. That is due to the exponential nature of the Fermi-Dirac distribution of electrons in both conduction and valence bands. As the cantilever vibrates, the chemical potential on the top side of the piezoelectric layer (attaching to the QDs) $\mu_P$ with respect to time $t$ is shown in the \textbf{Figure 3b}. Compared with the results in \textbf{Figure 2b}, $\mu_P$ curve displays a square wave instead of a sinusoidal, which coincides with the results of $\mu_e$ and $\mu_h$ versus carrier density. Further on, the electrochemical potential $\mu_P$ is used to calculate the equation (8) for the purpose of obtaining conductance as the function of time $t$. Conductance $G(t)$ is then obtained numerically for a single quantum dot. Suppose dots density is $2.78 \times 10^{14}$ $m^{-2}$, total conductance of the whole surface area of the cantilever $G_t$ can be considered to be the sum of many single conductances $G_0$ connected in parallel.  Then it is $G_t=\sum G_0= 6.94 \times 10^{8} G_0$.  The results are shown in the \textbf{Figure 4}, where both $R(t)$ (resistance) and conductance $G(t)$ curves are displayed. Conductance varies from a very low value ($1.74 \times 10^{-8} S$) to a high value ($0.07 S$), corresponding to resistance from $5.74 \times 10^7 \Omega$ to $14.5 \Omega$. A simple circuit consisting of the integrated device and a 1 $k\Omega$ resistor is sketched to demonstrate the rectification effect.  From \textbf{Figure 5}, clearly the rectification effect has been achieved and almost all positive part of the voltages has been blocked due to very large impedance of the QDs layer, attributed to its energy selection mechanism, where $V_1$ and $V_2$ refer to the equivalent circuit in \textbf{Figure 1c}. Graphical demonstration of conductance vs. chemical potential ($G$-$\mu_P$) curve with the varying $\delta z$ for a single QD is shown in \textbf{Figure 6}, which clearly displays a performance similar to that of diodes. As the current flows through the close-loop circuit consisting of a finite external load resistance, the electrochemical potential $\mu_P$ shifts to the neutral. Results from equation (5) (\textbf{Figure 3b}) reveals that the variation of the $\mu_P$  is mainly determined by the polarity of the accumulated charges in the QD-Piezoelectric interface. A small quantity of charges will bring the $\mu_P$ close to $\mu_c$ or $\mu_v$, and further increasing the quantity does not vary $\mu_P$ significantly (\textbf{Figure 3a}). This makes the metal-QD-piezoelectric interface as a perfect rectifier compared to the classic diodes that require a forward bias. Complex motions of the mechanical structure give charges with opposite signs on the the same surface, for example a 3rd resonant mode displayed in \textbf{Figure 7}. The mode shape is obtained from Euler$-$Bernoulli beam theory on cantilever resonance \cite{Timo}, the 3rd resonant mode is found by numerically solving the third root $\beta_3$ of the nonlinear equation $cosh(\beta_nL)cos(\beta_nL)+1=0$. $dz^2/dx^2$ is for the curvature at $x$ point of the cantilever. Subsequently the mode shape and charge distribution can be calculated. It is seen that both positive and negative charges are generated on the surface, causing charges cancellation. With the QDs layer, only one type of charges can flow through, reducing the current loss on electrode. Though this hypothesis is mainly based on the theoretical validation, experiments in previous published work \cite{nnqd} on thermoelectric energy devices showed that using QDs in power generation devices is viable. 

\section{Conclusion}

An integrated piezoelectric energy harvesting concept comprising a quantum dots layer and a piezoelectric layer stacked together is reported in this article. The new structure enables rectification function to be achieved without additional external circuits, as well as avoidance of the screening effect on the surface of the piezoelectric material. Multiphysical simulations taking account of the mechanical-electrical, charge-chemical potential, and elastic tunnelling process validate the concept. The modelling results show clearly that the rectification effect has been achieved. It is worth mentioning that this configuration can be applied to other structures besides the cantilever shape.

\section{Acknowledgements}

LL appreciates the support of college of engineering, Swansea
University.

\section{Additional information}

The author declares no competing financial interests.

\bibliography{refs}

\newpage

\begin{figure}
\centerline{\includegraphics[height=13.cm]{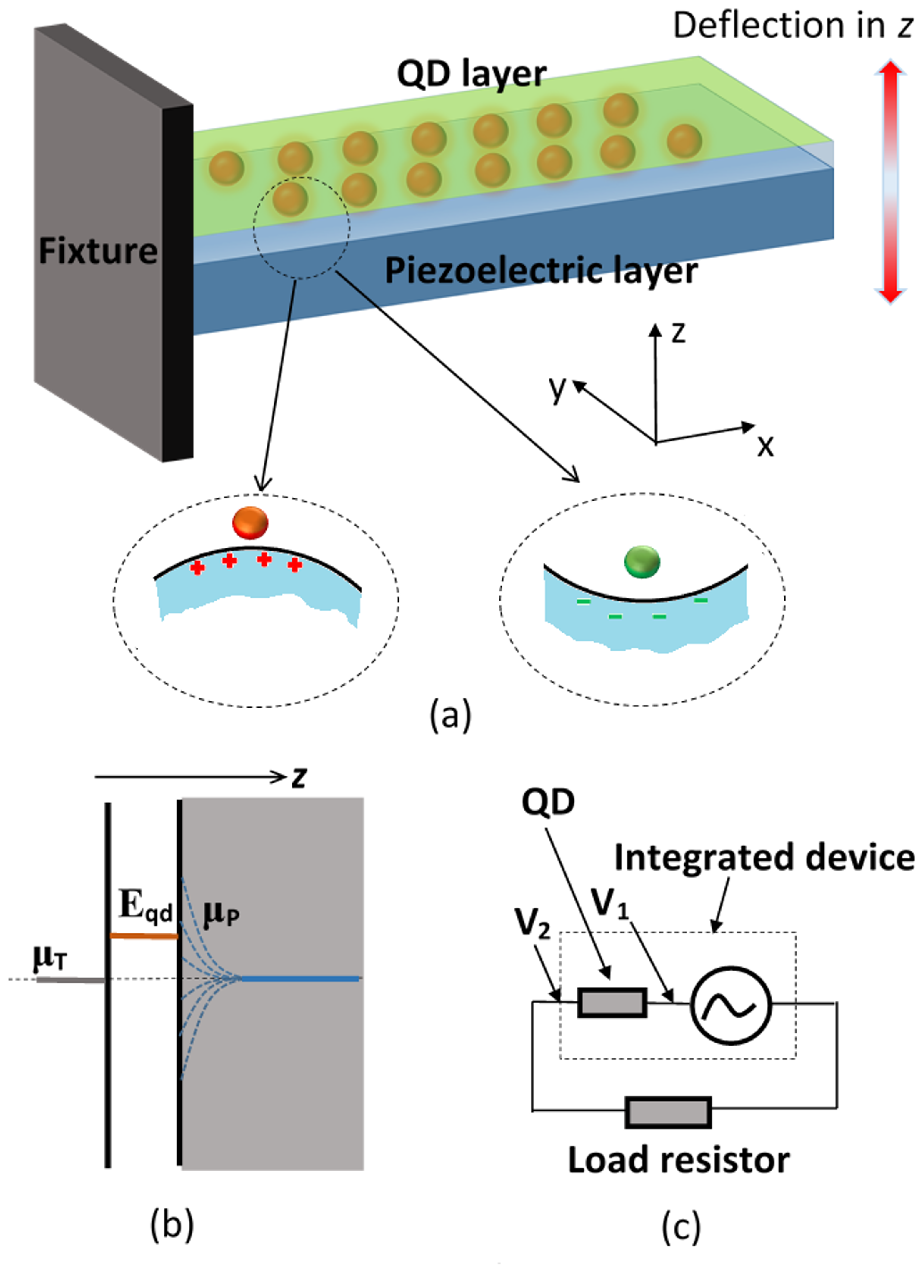}}
  \caption{Schematic view of the integrated energy harvester. (a) 3D view of the stacked QD and piezoelectric layers. Circled diagrams show the principle of adaptable conductance of the QD subjecting to different surface charges, where red dot represents low conductance, green dot shows high conductance. (b) Energy diagram of the electrode-QD-piezoelectric interface. (c) equivalent circuit of the device.} 
    \label{fig_1}
\end{figure}

\begin{figure}
\centerline{\includegraphics[height=9 cm]{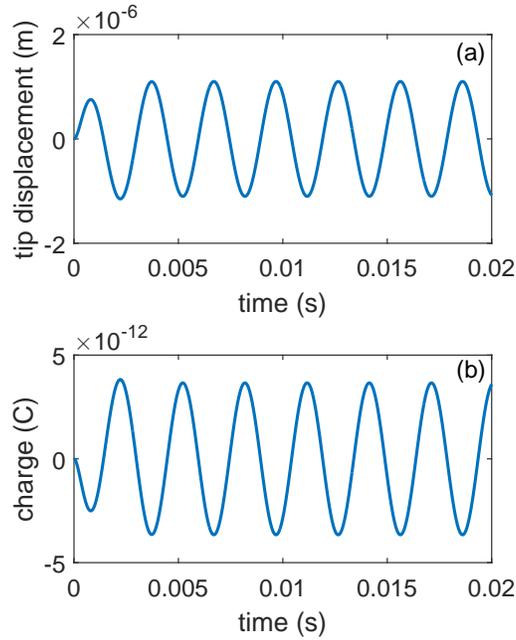}}
  \caption{Vibration of the cantilever inducing electrical charges, frequency is calculated to be 335.9 Hz. (a) Tip deflection vs. time. (b) Generated charge vs. time $t$.} 
    \label{fig_2}
\end{figure}

\begin{figure}
\centerline{\includegraphics[height=10 cm]{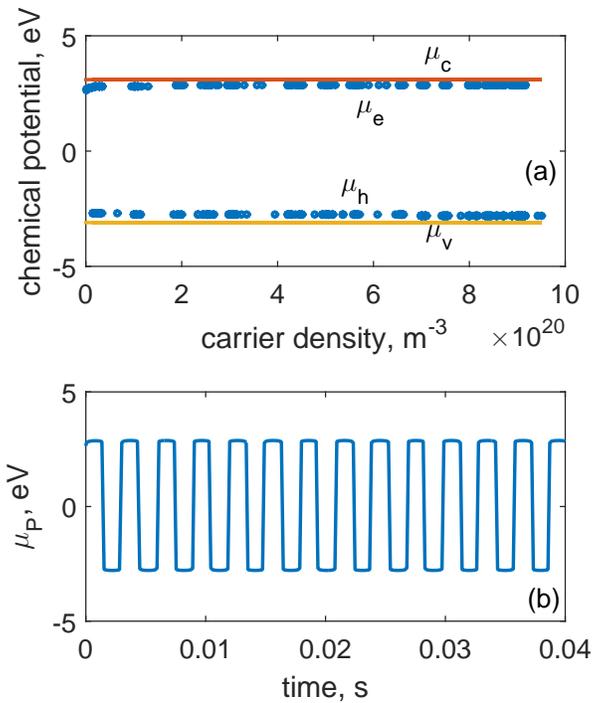}}
  \caption{Chemical potentials calculated from the increase of carrier density and polarity change of generated charges on the piezoelectric surface due to vibration of the piezoelectric cantilever. (a) $\mu_e$, $\mu_h$, $\mu_c$, and $\mu_v$ in relation to the carrier density. (b) $\mu_P$ vs. time $t$.} 
    \label{fig_3}
\end{figure}

\begin{figure}
\centerline{\includegraphics[height=8.cm]{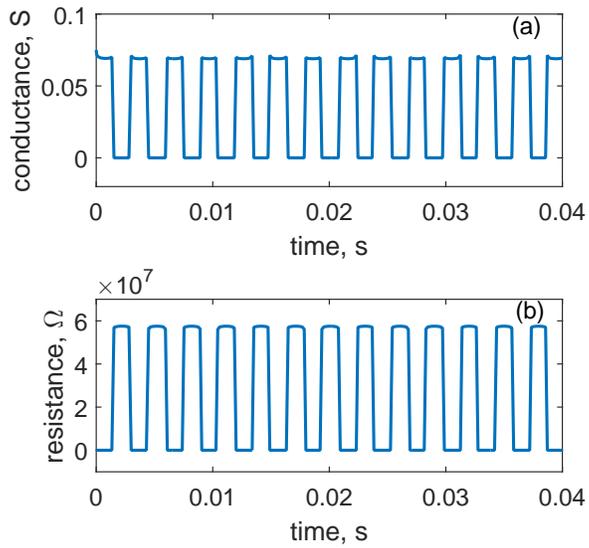}}
  \caption{Calculated conductance and resistance of the QD layer with respect to time. (a) Conductance vs. $t$. (b) Resistance vs. $t$.} 
    \label{fig_4}
\end{figure}

\begin{figure}
\centerline{\includegraphics[height=8.5cm]{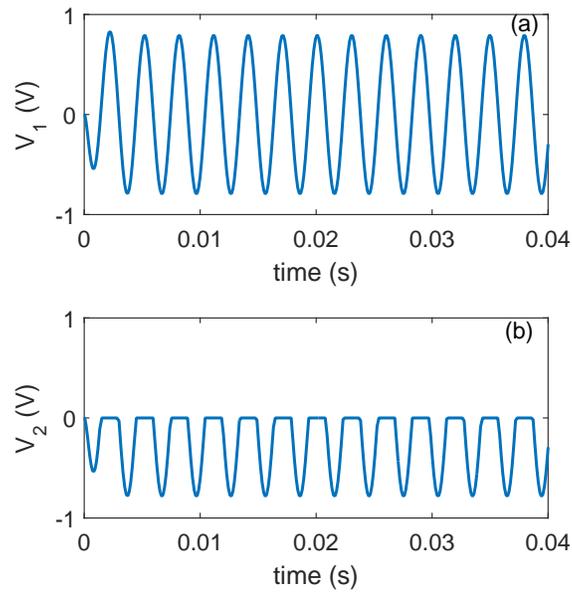}}
  \caption{Simulation of the rectification effect of the integrated piezoelectric harvester. (a) Voltage with respect to the time for piezoelectric layer. (b) Voltage vs. $t$ for the integrated device} 
    \label{fig_5}
\end{figure}

\begin{figure}
\centerline{\includegraphics[height=7.cm]{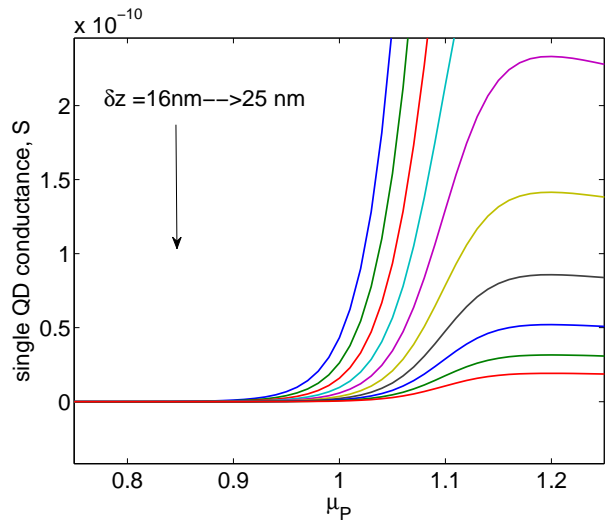}}
  \caption{Conductance vs. chemical potential of piezoelectric layer} 
    \label{fig_6}
\end{figure}

\begin{figure}
\centerline{\includegraphics[height=8.cm]{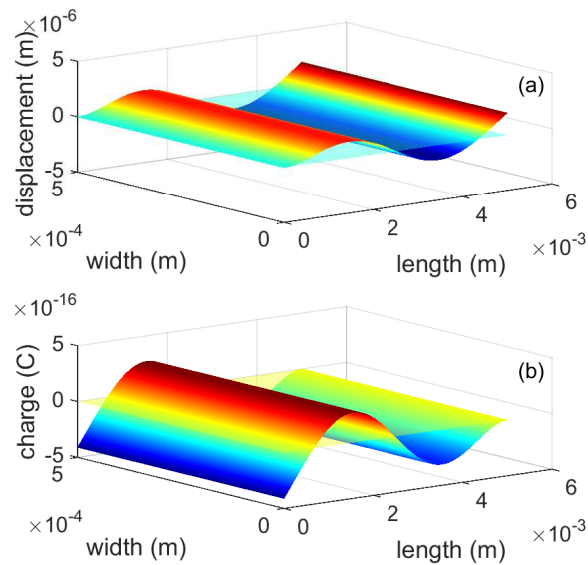}}
  \caption{Third resonant mode shape of the cantilever and charge distribution. (a) 3D view of the mode shape. (b) Charge distribution. Without QDs layer, the net value of charges is $-2.6 \times 10^{-11}$C. With QDs layer, the net value of charges is $-5.8 \times 10^{-11}$C. Cantilever in static and charge in neutral are also shown in the figure (semi-transparent plane).} 
    \label{fig_7}
\end{figure}

\end{document}